\documentclass[preprint]{imsart}

\usepackage{amsthm,amsmath,natbib}
\RequirePackage[colorlinks,citecolor=blue,urlcolor=blue]{hyperref}
\usepackage{graphicx}
\startlocaldefs
\endlocaldefs

\begin{document}

\begin{frontmatter}

\title{New estimates for network sampling}
\runtitle{Estimates for network sampling}


\author{\fnms{Steve}
  \snm{Thompson}\corref{}\ead[label=e1]{thompson@sfu.ca}}
\address{Department of Statistics and Actuarial Science\\8888
  University Drive\\Burnaby, BC V5A 1S6 Canada\\\printead{e1}}
\affiliation{Simon Fraser University}

\runauthor{Steve Thompson}

\begin{abstract}
  Network sampling is used around the world for surveys of vulnerable,
  hard-to-reach populations including people at risk for HIV, opioid
  misuse, and emerging epidemics.  The sampling methods include
  tracing social links to add new people to the sample.  Current
  estimates from these surveys are inaccurate, with large biases and
  mean squared errors and unreliable confidence intervals.  New
  estimators are introduced here which eliminate almost all of the
  bias, have much lower mean squared error, and enable confidence
  intervals with good properties.  The improvement is attained by
  avoiding unrealistic assumptions about the population network and
  the design, instead using the topology of the sample network data
  together with the sampling design actually used.  In simulations
  using the real network of an at-risk population, the new estimates
  eliminate almost all the bias and have mean squared-errors that are
  2 to 92 times lower than those of current estimators.  The new
  estimators are effective with a wide variety of network designs
  including those with strongly restricted branching such as
  Respondent-Driven Sampling and freely branching designs such as
  Snowball Sampling.
\end{abstract}


\end{frontmatter}


\section{Introduction}

\begin{figure}
\centering
\includegraphics[width=1.0\linewidth, height=1.05\linewidth]{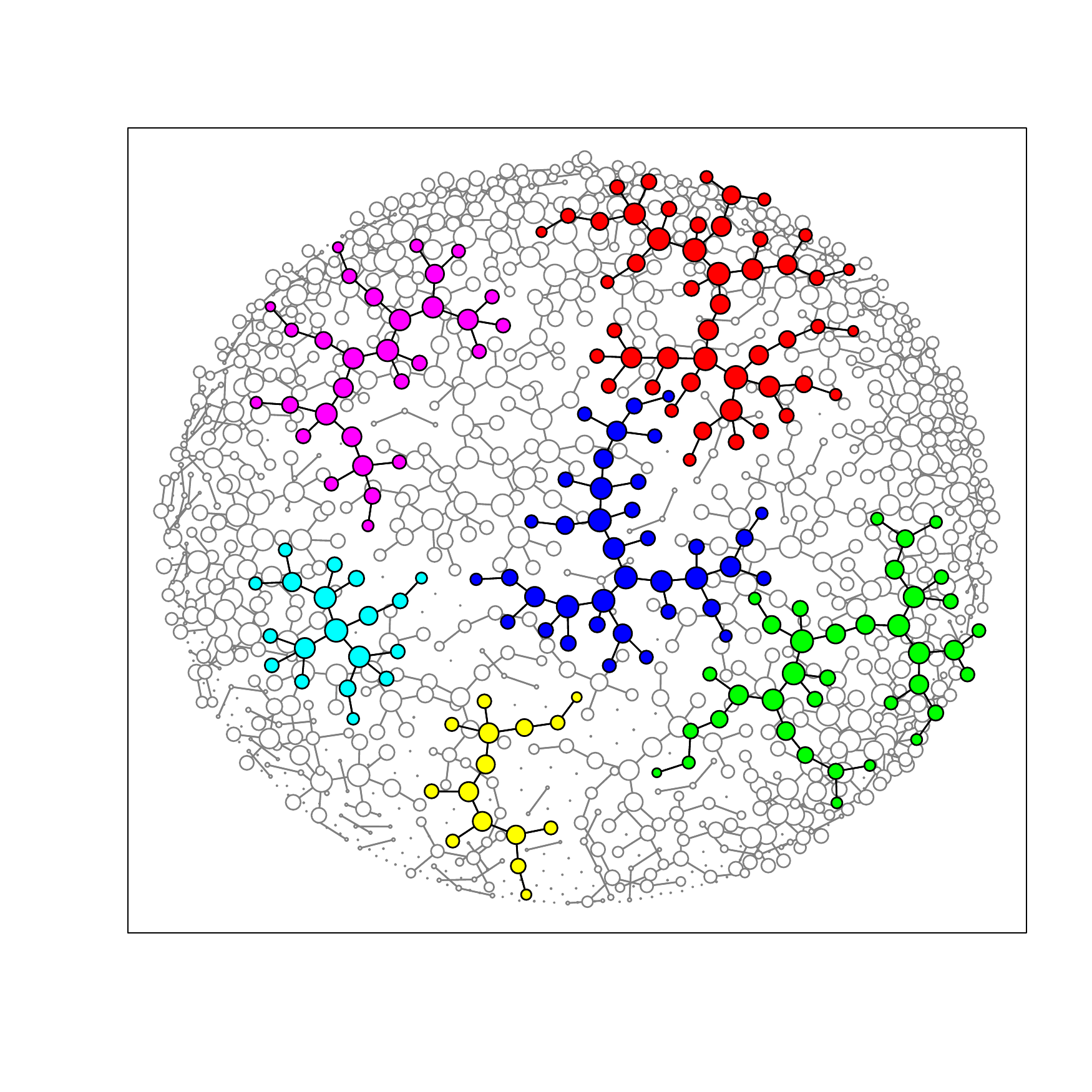}
\caption{A network sample of 1200 people from a high-risk hidden
  population. Some sample components are highlighted.  The network
  sample is resampled many times using a network sampling design that
  adheres closely to the original survey design.  The frequency that a
  person is included in the resamples (circle diameter) is used to
  estimate the person's inclusion probability in the original sample.
  These estimated inclusion probabilities enable estimates of
  hidden-population characteristics.  }
\label{fig:sample}
\end{figure}

Network sample surveys are in wide use around the world for studies of
hard-to-reach vulnerable populations [\cite{white2015strengthening},
  \cite{verdery2015network}].  In these surveys social links are
followed from people already in the sample to find and bring new
people into the sample. For the key populations at high risk for HIV
these methods provide the most effective way to gain scientific
understanding about the behavioral, biological, and network risks.
These studies have been supported by the U.S. President's Emergency
Plan for AIDS Relief (PEPFAR) and the Centers for Disease Control and
Prevention (CDC) and other national and international organizations,
with research on the methodologies supported by the National Science
Foundation and National Institutes of Health NIH
[\cite{mouw2012network}].  The annual UNAIDS statistics on HIV depend in
part on network surveys [\cite{unaids2018report}].  The surveys are
essential for basic scientific understanding, for assessing network
risk even before any virus moves in, and for evaluating the
effectiveness of intervention programs.  Policy decisions in response
to outbreaks of HIV and other emerging epidemics require accurate
estimates from these network surveys.

The estimation methods currently used with the survey data are known
to be inaccurate, with large mean squared errors and biases and
unreliable confidence intervals [\cite{goel2010assessing},
  \cite{gile20107}].  The inaccuracies arise because different people
in the hidden population have different probabilities of coming into
the sample.  A person in a highly connected region of the population
will have a higher inclusion probability than one in a less connected
position.  The adjustments currently made for these unequal inclusion
probabilities are based on assumptions about the sampling design that
are discrepant from the survey designs actually used. The estimators
derived from the unrealistic assumptions do not use the full sample
data available and in particular ignore the network topology of the
sample.

New estimators introduced in this report remove almost all of the
bias, have much lower mean squared errors, and enable confidence
intervals having good coverage probabilities and modest widths.  The
improvements are achieved by using the full sample network data and
incorporating the key features of the actual survey design.  For the
new estimates, a sampling process similar to the actual survey design
is run on the sample network data.  The frequency with which a person
is included in the sampling process provides an estimate of that
person's inclusion probability in the actual survey.  The estimated
inclusion probabilities are then used in a generalized unequal
probability estimator to estimate characteristics of the hidden
population, such as proportion infected with a virus, prevalence of a
risk behavior such as exchanging sex for money, mean number of
partners per person, or rate of concurrent relationships.

In the actual network survey, the probability that a person is
included depends on the person's position in the social network
topology in interaction with the type of sampling design used.  The new
method estimates the inclusion probabilities from the sample network
topology in interaction with the same type of sampling design.

Figure 1 shows the network topology of a sample of 1200 people from a
high-risk population of drug users, sex workers, clients and other
partners.  The nodes represent individuals (circles) and the lines
represent sexual, drug, and social relationships between pairs of
individuals.  The sample network topology emerges as the connectedness
pattern of the paths.  The sample network topology includes separate
sample components, several of which are highlighted.  Within a
component, the sample network topology has a tree structure.  The tree
structure is a result of the design protocol, in which a person is not
allowed to be recruited more than once.  The population network, from
which the sample is selected, is not a tree but has a more general
net-like topology and larger components than appear in the sample.

The sample of Figure 1 was selected by the most commonly used form of
network sampling, Respondent-Driven-Sampling (RDS), in which an
initial sample of seeds is selected by investigators and each person
is given up to three coupons with which to recruit up to three of
their partners into the sample. When a person with a coupon comes in
to be interviewed, that recruit is in turn given three coupons which
which to recruit additional sample members, and so on.  An individual
is not allowed to be recruited more than once, which results in the
tree structure of the sample topology.  The three-coupon limit
restricts branching of the network sample to three.  So counting the
person who recruited the individual gives a maximum of four links from
any individual in Figure 1.  Not every coupon issued is used, and not
every individual recruited comes in to be interviewed, although there
is a monetary incentive offered to both the recruitee and the
recruiter.

If unlimited branching is allowed, or if a high coupon limit such as
15 is used, the network design is typically referred to as Snowball
(SB) sampling.  Snowball samples tend to include some larger sample
components.  With either design, coupons are given an expiration date,
such as 28 days from the date of issue.  Recruitment in a sample
component comes to an end if there are no links out from people in the
sample with unexpired coupons to people not already in the sample.

Early uses of network sampling for hard-to-reach populations 
typically used Snowball Sampling methods, with survey data summarized
by unweighted sample means and proportions [\cite{spreen1992rare},
\cite{heckathorn1997respondent}, and \cite{thompson2002adaptive}].
Statistical estimates of population values from relatively simple
network sampling designs were obtained by \cite{birnbaum1965design}
and \cite{frank1977survey}, \cite{frank1994estimating}.

Starting with \cite{heckathorn1997respondent}, the methodology of
Respondent-Driven-Sampling using dual-incentive coupons was
introduced.  Estimators for these designs based on random walk theory
and assumptions of Markov transitions in the sampling between values
of attribute variables of respondents were given in
\cite{salganik2004sampling}, \cite{heckathorn20076}, and
\cite{volz2008probability}. If a random walk with replacement is run
in a network consisting of a single connected component, the long term
frequency of inclusion of node $i$ is proportional to $d_i$.  One
consequence of the use of random-walk theory to justify the estimators
was that the more freely branching snowball designs became more seldomly
used, with the idea that the designs restricting branching to 3 or
fewer links would be closer to the assumed random walk model.

The estimator of \cite{volz2008probability} (VH estimator), uses $d_i$
in place of actual inclusion probability in an unequal probability
estimator.  The estimator of \cite{salganik2004sampling} (SH
estimator) used the same form as VH to estimate mean degree, and for
means of binary attribute variables adjusts that with proportions of
sample recruitment links between and within the group with the
attribute and the group without it.  The adjustment is based on the
additional assumption of Markov transitions between attribute states
during the sampling.  Adjustments for surveys in which sample size is
a large fraction of population size include the Successive Sampling
(SS) estimator \cite{gile2011improved} for the VH estimator and the
Homophily Configuration Graph (HCG) estimator \cite{fellows2018respondent} for
the SH estimator.

A random walk design in a network at each step allows tracing of only
one randomly selected link from the current node (person), and
sampling is done with replacement so the same person can be selected
again later.  The inaccuracies of the widely-used current estimators
result from the discrepancy between the assumed random walk model,
which has no branching and is with-replacement, and the designs
actually used in the surveys, with their branching and
without-replacement sampling.  Additionally, the current estimators
use an assumption that the population network has only a single
component, in support of the theoretical properties of the random walk
model.  Evidence suggests that many real population include more than
one connected components.  An additional assumption underlying some of
the current approaches to estimation assumes first-order Markov chain
transitions during the sampling between attribute values of selected
nodes.  The types of networks that would support this assumption are
unlikely to be encountered with real network populations
[\cite{verdery2015network}].

Currently used estimators such as the VH and related estimators do not
make use of the sample network topology of the data.  Instead, they
use the number of partners (degree) reported by each individual.  The
SH and related estimators use, in addition to degree, counts of sample
links between groups having different values of an attribute, but not
the pattern of network paths by which the links connect together.

Confidence interval methods proposed for RDS designs have most often
been based on bootstrap methods [ \cite{salganik2006variance},
\cite{gile2011improved}].  Evaluations of these methods with
various real and simulated network populations include
\cite{spiller2017evaluating} and \cite{baraff2016estimating}.

The concern of the work above and of the present report is estimation
of population characteristics, such as prevalence of infection,
proportion of people with a risk-related behavior, mean degree, or
concurrency.  The related problem of estimating the size of a hidden
population with network sampling is addressed in
\cite{handcock2014estimating}, \cite{crawford2016graphical},
\cite{crawford2018hidden}, and \cite{vincent2017estimating}. Another
closely related problem is use of network designs to adaptively spread
interventions in a population [\cite{valente2012network}].  In fact
network sampling for at-risk populations as described in this report
whose primary purpose is fundamental knowledge and estimation usually
bring beneficial interventions as well to the hard-to-reach
population.  Such interventions include testing for HIV and other
infections, referral to medical services and enhanced adherence
counselling for individuals who test positive, referral to addiction
treatment programs, and distribution of condoms or clean injection
equipment.  Link-tracing designs have been shown to be a highly
effective way to introduce and distribute interventions in a
population [\cite{thompson2017adaptive}].

 An RDS design to study a rural opioid user network in relation to HIV
 risk was used in \cite{young2018network}, \cite{young2014spatial},
 and \cite{young2013network}. The project used an additional design
 feature of a social network study of sample members which enabled
 examination of differential recruitment rates for partners having
 different attribute values.

For the new estimation approach described here we make no assumption
of any form or model for the population network.  Instead we use the
sample network and its topology just as it is in the data.  Instead of
any unrealistic assumption about the sampling design, we use the same
type of design as used in the actual survey, and use it to select many
re-samples of the sample network.  The actual survey design for a
survey is documented in the survey protocol document and survey
reports.  Typically it allows branching up to the coupon limit and is
done without replacement.  

The theory used for the new estimation method is the design-based
theory of inference in sampling [\cite{sarndal1978design}], in which the
population---here including its network---is conceived as fixed but
unknown and inference is done using the design-induced probabilities
by which the sample was selected.  The key to the new method is to use
the full sample network to estimate the probability with which each
person was included in the sample.  These inclusion probabilities can
no be known exactly, because they depend on the population network
topology both inside the sample and outside it.

The idea of the new method is very simple (first described in the
technical papers \cite{thompson2018simple} and
\cite{thompson2019design}).  We run a sampling process similar to the
actual survey design on the sample network data and use the inclusion
frequencies in the sampling process to estimate the inclusion
probabilities in the real-world sampling.  With the estimated
inclusion probabilities, well-established sampling inference methods
can be used for population estimates and confidence intervals. Details
of the re-sampling and estimation are described in the Supplement.
Two approaches to the re-sampling are repeated re-samples and a Markov
chain resampling process.  For the simulations we use the resampling
process because it is computationally so fast.

The size of a node in Figure 1 is drawn with diameter proportional to
the inclusion frequency of that person in the re-sampling process.  A
node in a more central position in a component has a higher inclusion
frequency than one is a more peripheral position in the sample network
topology.  Nodes in larger components also tend to have higher
inclusion frequencies because more network paths lead to them.  If we
look at nodes with just one link in Figure 1, they vary in size one to
another, with those connected to a more central part of the component
having higher inclusion frequency.  A similar variation in size can be
seen among nodes with two links, and so on.  So inclusion frequency is
not a simple function of node degree but depends on position in the
network topology.  Note that the relevant centrality measured here by
node diameter is not a calculation from the network topology alone but
with the interaction of that topology with the branching network
sampling design.

\section{Methods}

\subsection*{The new estimation method}

In traditional survey sampling with unequal probabilities of inclusion
for different people, typical estimators divide an observed value
$y_i$ for the $i$th person by the inclusion probability $\pi_i$ that
person.  A variable of interest $y_i$ can be binary, for example 1 if
the person tests positive for a virus and 0 otherwise, or can be more
generally quantitative, such as viral load.  The inverse-weighting
gives an unbiased or low-bias estimate of the population proportion or
mean of that variable.  In network surveys the inclusion probabilities
are unknown so they need to be estimated.

The estimators described in this report first estimate the inclusion
probability of each person in the sample by selecting many resamples
from the network sample data using a design that adheres in key
features to the actual survey design used to find the sample.  In
particular, the resampling design is a link-tracing design done
without-replacement and with branching, as was the original design.
The frequency $f_i$ with which an individual is included in the
resamples is used as an estimate of that person's inclusion
probability $\pi_i$.

What we do is select $T$ resamples $S_1, S_2, ..., S_T$ from the
sample network data.  There are two approaches to selecting the
sequence of resamples.  In the repeated-samples approach each resample
is selected independently from seeds and progresses step-by-step to
target resample size independently of every other resample, so we get
a collection of independent resamples.  in the sampling-process
approach each resample $S_t$ is selected from the resample $S_{t-1}$
just before it by randomly tracing a few links out and randomly
removing a few nodes from the previous resample and using a small rate
of re-seeding so we do not get locked out of any component by chance.
It is this Markov resampling process approach that we use for the
simulations in this paper because it is so computationally efficient.

For an individual $i$ in the original sample, there is a sequence of
zeros and ones $Z_{i1}, Z_{i2}, ..., Z_{iT}$, where $Z_t$ is 1 if that
person is included in resample $S_t$ and is 0 if the person is not
included in that resample.  The inclusion frequency for person $i$ is 
\begin{equation}
f_i =\frac{1}{T}\left( Z_{i1} + Z_{i2} + ... + Z_{iT} \right)
\label{eq:fi}
\end{equation}

In Figure 1 the circle representing individual $i$ in
the sample is drawn with diameter proportional to the estimated
inclusion probability $f_i$ of that individual.  Individuals centrally
located in sample components tend to have high values of $f_i$.  That
is because there are more paths, and paths of higher probability,
leading the sample to those individuals.  Also, individuals in larger
components tend to have larger $f_i$ than individuals in smaller
components, so that the method is estimating inclusion probability of
an individual relative to all other sample units, not just those in
the same component or local area or the sample This is because of the
self-allocation of the branching design, even in the absence of
re-seeding, to areas of the social network having more links and
connected paths.

The estimator of the mean of a characteristic $y$ in the hidden
population is then 
\begin{equation}
\hat\mu_f = \frac{\sum (y_i/f_i)}{\sum (1/f_i)}
\label{eq:est}
\end{equation}
where each sum is over all the people in the sample.  If the actual
inclusion probabilities $\pi_i$ were known and replaced the $f_i$ in
Equation \ref{eq:est} we would have the generalized unequal
probability estimator $\hat\mu_\pi$ of Brewer \cite{brewer1963ratio}.

\subsection*{Why it works}

To understand why the new method works, consider the two stages of
sampling.  The first-stage is the actual network sampling design by
which the sample of people is selected from the hidden population.
The second-stage design selects a resample of people from the network
sample data, using a network sampling design similar to the one used
in the real-world.  The second-stage design, like the first, uses
link-tracing, branches, and is done without-replacement.  The
second-stage design can not be exactly the same as the original design
in every respect.  For example the second-stage design has to use a
smaller sample size than the original, because of the
without-replacement sampling.

The probability that individual $i$ in the
original sample is included in the resample will be called $\phi_i$.  
The ideal is to have the inclusion probability for
a unit at the second stage, given the first stage sample, to be
proportional to it's inclusion probability in the original design.
That is, $\phi_i = c\pi_i$, where $c$ is some constant, which does not
need to be known.

Now let $\hat\mu_{\phi}$ be formula \ref{eq:est} with the exact
resample probabilities $\phi_i$ replacing $f_i$.  If $\phi_i = c\pi_i$,
then $\hat\mu_{\phi} = \hat \mu_{\pi}$, because the constant of
proportionality $c$ is in both the numerator and denominator of
\ref{eq:est} and divides out of the estimator.

  As the number of resamples $T$ gets large the inclusion
frequencies $f_i$ converge in probability to the second-stage
inclusion probabilities $\phi_i$.  This is by the (weak) Law of Large
Numbers for the independent resamples and by the Law of Large Numbers
for Markov chains for the resampling process that traces a few and
removes a few at each step.

It follows that if $\phi_i$ is proportional to $\pi_i$ then
$\hat\mu_f$ converges in probability to $\hat\mu_\phi$.  So if
inclusion in the resample $\phi$ is proportional to inclusion in the
original sample $\pi$ then the estimator we use here $\hat\mu_f$
converges to the general unequal probability estimator
$\hat\mu_{\pi}$.  Since the resampling is fast computationally,
especially with the sampling process approach, we can readily select a
lot of resamples, such as $T = 10,000$ that we use in the simulations
here, and higher values of $T$ like one million are still fast to
compute.

The approximate part is in how close the second-stage inclusion
probabilities $\phi_i$ are to proportionality with the first-stage
inclusion probabilities $\phi_i$.  This is why it is important that
the resampling design adheres to the main features of the actual
network design, such as network link tracing, branching, and
without-replacement selections.

The use of the second stage sample is different here than in
traditional two-stage sampling or in bootstrap methods.  In each of
those a given second-stage sample is used to make an estimate of a
population value. In the case of bootstrap methods, many such estimates, from the many resamples, are used to construct a confidence interval. If the sampling is with unequal probabilities at
each stage, estimation of the population value from the second-stage
sample requires dividing first by the second-stage inclusion
probability $\phi_i$ to estimate what is in the first stage sample and
then then by the first stage probability $\pi_i$.  Here we use the
second stage design solely to estimate its own inclusion probabilities
and we construct the second-stage design to have those probabilities
similar to the first-stage probabilities.

\subsection*{Confidence intervals}

When an estimator has a large bias, it is hard to have a confidence
interval with adequate coverage probability based on that type of
estimator.  Such an interval has to be extra wide to accommodate not
only the sampling variance but the offset between the expected value
of the estimator and the true population value. This is why confidence
intervals for the current estimators have been problematical.  The
confidence interval here takes advantage of the better estimate we get
by using the resampling process inclusion frequencies $fi$ to estimate
the actual inclusion probabilities.  The confidence interval method
for the new estimators uses the $fi$ again in forming the interval,
using unequal probability sampling methods.

A simple variance estimator to go with $\hat\mu_f$ is
\begin{equation}
\widehat{\textrm var}(\hat\mu) =  \frac{1}{n(n-1)}  \sum_{s}
\left(\frac{ny_i/f_i}{\sum_s (1/f_i)} - \hat\mu_f\right)^2
\end{equation}
This is based on a variance estimator from unequal probability
sampling but is modified here to serve the generalized unequal
probability estimator and uses the inclusion frequencies $f_i$ in
place of the unknown actual inclusion probabilities.  An approximate
$1 - \alpha$ confidence interval is then calculated as $ \hat\mu \pm z
\sqrt{\widehat{\textrm var}(\hat\mu)}$, with $z$ the $1-\alpha/2$
quantile from the standard Normal distribution.  In the simulations
here the estimator above gives a slightly higher average confidence
interval coverage probability.

\section{Results}

The new estimators are evaluated and compared with the current
estimates using the network data on the hidden population at risk for
HIV enumerated in the Colorado Springs study on the heteroxexual
transmission of HIV [\cite{potterat1999network}], also known as the
Project 90 study.  The study was so thorough in finding every linked
person that it provides the most relevant network data set that can be
considered as an entire at-risk population for the purpose of
comparing sampling designs and estimators.  The population and the
simulation methods are described in more detail in the Methods
section.

The most commonly used estimator with network surveys in current
practice is the VH estimator.  The other variations in use such as SH
are related to it and are based on the same assummptions plus the
additional assuption of a first-order Markov process in transitions
between node attribute values during the sampling.  The SH estimator
is used mainly for binary attribute variables.  For categorical
variables it has the property that the proportion estimates do not add
to one without extra adjusting of one kind or another, and it is not
well suited to continuous variables.  \cite{goel2010assessing} found
that the SH estimator performed about the same as the VH estimator,
and \cite{gile20107} found SH performed a little less well than the VH
estimator.    For all of these reasons we use
the VH estimator as the basis of comparisons with the new estimator in
this section.

\begin{figure}
\centering
\includegraphics[width=0.9\linewidth]{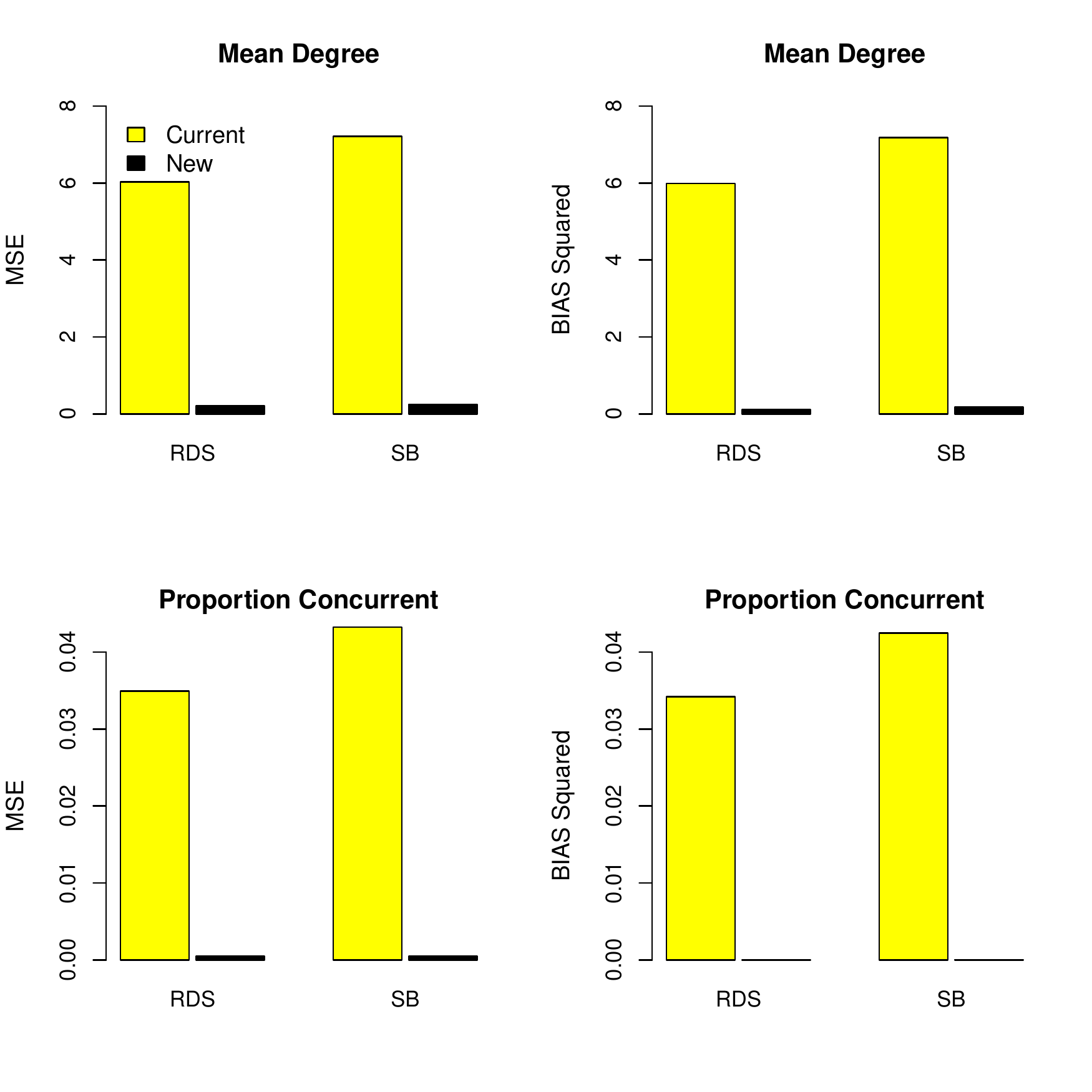}
\caption{The mean square error (left) and bias (right) of the new
  estimator (black) is lower than that of the current estimator
  (yellow) for estimating mean degree (top) and 
  concurrency---proportion of people with two or more partners
  (bottom) for each  network sampling design.  The design RDS
  restricts branching by limiting coupons to a maximum of 3.  The
  snowball (SB) design gives respondents as many coupons as their
  number of partners, up to a maximum of 15, allowing almost unlimited
  branching as the sample is selected.  }
\label{fig:bar}
\end{figure}

\subsection*{Link-related variables}

Among the most important quantities to estimate in relation to spread
of HIV are the means and proportions of link-related variables.  Two
of widespread interest are mean degree) and concurrency.  Mean degree
is the average number of partners per person in the population.  The
most common definition oncurrency is the proportion of people in the
population who have two or more partners.  This and related
definitions of concurrency and their role in spreading HIV are
discussed in \cite{kretzschmar1996measures}, \cite{morris1997concurrent}, and
\cite{admiraal2016modeling}.  A high number for either of these is an
indication that an epidemic could spread rapidly in the population
once it starts there.

The top two plots in Figure \ref{fig:bar} are on estimating mean
degree.  Plots on the left show MSE.  Plots on the right show squared
bias.  The top row is for the RDS design with 3 coupons.  The bottom
row is for the SB design with 15 coupons. The current estimator (VH)
is shown in yellow and the new estimator in black.  For any estimator,
the MSE equals the variance plus the bias squared.  MSE and bias
squared are plotted on the same scale, so it is easy to see that most
of the reduction of bias with the new estimator comes by reducing,
almost eliminating, the bias.

Starting with the left-most pair of bars, the mean square error for
the current method for RDS sampling is 6.03.  The mean square error
for the new estimator is 0.21.  The reduction in mean squared error
comes largely from eliminating most of the bias, as shown in the top
right plot of Figure 2.  The actual mean degree in the Colorado Springs
study population is 7.89 partners per person.  The current estimator
underestimates this on average by 2.45 partners.  The new
estimator overestimates by only 0.32 on average.

 Here the squared bias 5.99 accounts for almost all of the MSE 6.03.
 The new estimator reduces the squared bias to 0.11.  The reduction in
 bias is obtained by the more accurate estimates of inclusion
 probabilities using a the resampling design that adheres to the the
 branching and without-replacement features of the actual survey
 design, and using the sample network recruitment data instead of
 assumptions about the hidden population network.

For estimating mean degree, with either the RDS or the SB design the
relative efficiency (ratio of MSEs) of the new estimator is 29.  So
with the same design, the new estimator reduces the mean squared error
to 1/29 that of the current method.  For estimating the proportion of
people in the population with two or more partners (concurrency,
bottom row in Figure 2), the relative efficiencies of the new
estimators compared to the current estimators are 72 for RDS and 92
for the SB design.

The reason for the dramatic improvement of the new estimates over the
current estimates for the link-related quantities is that a
link-related variable such as number of partners or having two or more
partners is strongly related to the actual inclusion probabilities of
the link-tracing design, whether the design is RDS or SB.  Therefore
with these variables it helps very much to have accurate estimates of
the inclusion probabilities, and large deviations from those probabilities
result in poor estimator performancce.  

\begin{figure}
\centering
\includegraphics[width=1.0\linewidth]{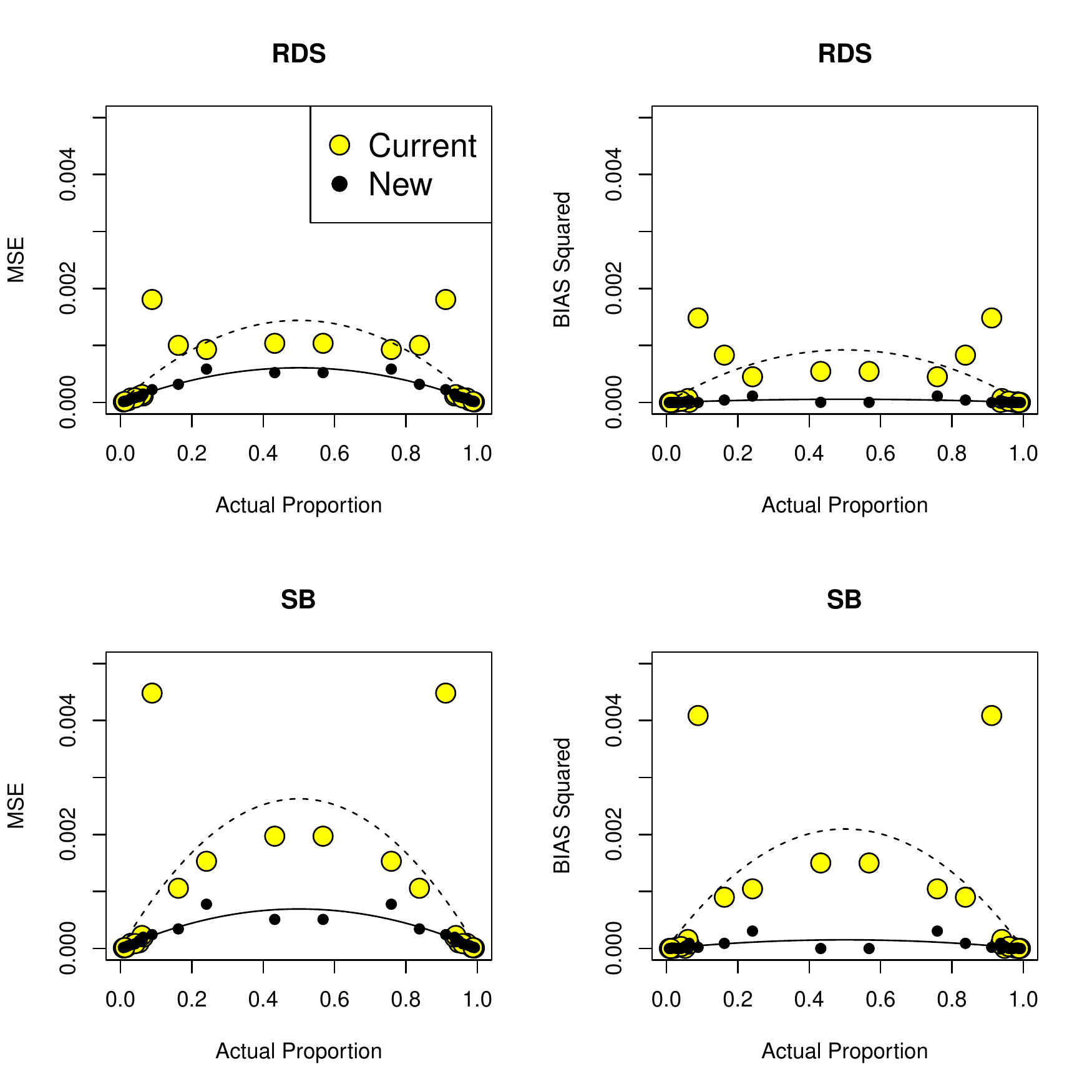}
\caption{Mean squared error of the estimate of population proportion
  for each of 13 individual attributes. The new estimator (black)
  is compared to the current estimator (yellow).  To help see the
  pattern, the estimate of the compliment of each attribute is shown.
  The compliment of sex work client, for example, is not-client.  A
  parabolic curve is fitted by weighted least squares to the MSEs of
  the new estimator (solid line) and the current estimator
  (dashed line).  The new estimator MSEs have a lower fitted
  curve with a better fit with each design.  }
\label{fig:parabola}
\end{figure}

\subsection*{Node attribute variables}

The Colorado Springs Study node data includes 13 individual attribute variables
such as sex worker, client of sex worker, or unemployed.  These are
variables 2 through 14 in Table 1.  For an individual, the value is 1
if the individual has the attribute and 0 otherwise.  The object for
inference for each attribute is to estimate the proportion of people
in the population having that attribute.  Most of these variables,
such as sex or race or employment status, are not strongly or
consistently related to inclusion probabilities.  Still, for
design-based estimators to work well it helps to have the estimated
inclusion probabilities close to the actual inclusion probabilities.
For the simulations, missing values were arbitrarily set to zero so
that sample sizes would be the same for all variables.


With a conventional simple random sampling design for estimating a
population proportion, the mean squared error of the estimate is a
parabola-shaped function of the actual population proportion.  The
actual proportion has to be between zero and one.  The MSE is highest
when the actual proportion is one-half and the MSE is zero when the
actual proportion is zero or one.  With the network designs and their
unequal inclusion probabilities the situation is more complex, but it
is still the case that the actual proportion has to be between zero
and one and that the MSE will be zero if the actual proportion is zero
or one.

To help see the pattern in the mean squared errors for estimating the
population proportions of the 13 attribute variables, the MSE for each
variable is plotted in Figure 3 against the actual proportion of
people having that attribute in the Colorado Springs study population.  For
each of the 13 variables, we can also estimate the proportion for its
complement.  The compliment of ``client'', for example, is ``not
client''.  The proportion for the compliment is 1 minus the proportion
with the attribute, and the MSE for estimation the compliment is the
same as the the MSE for estimation of the original variable.  This
gives us 28 variables for each plot in Figure 3, with actual
proportions ranging from near 0 to near 1.  The original variables are
on the left, since the actual proportions are all less than one-half.
The compliment variables provide redundant information but clarify the
pattern in the MSEs.

The MSE with the new estimator (black in plots) is lower than that 
of the current estimator in all cases except for some of the ones with
actual proportion near to zero for which the MSE is very small with
either estimator.  While the MSEs of the new estimator fall
rather close to the fitted parabola (solid line), the MSEs of the
current estimator are more erratic and the fitted parabola (solid
line) is higher.  The overall higher MSEs and erratic pattern with the
current estimator result from the discrepancies between actual
inclusion probabilities and those used in estimation.   

The parabolas in the plots have form ${\rm MSE}=ap(1-p)$ where $p$ is the
actual proportion, which is known for each of the 13 attribute
variables in the simulation population. The coefficient $a$ measures
the height of the fitted parabola for a given estimator-design
combination.  Since the relationship is linear with increasing variance in the
quantity $p(1-p)$, the weighted least squares regression estimator of
the coefficient $a$ is a simple ratio estimator.  The parabola height
provides a useful summary of the overall performance of the
estimator-design combination.  Because of the simple nature of a ratio
estimate, the ratio of height coefficients of two estimator-design
combinations is simply the the average MSE for the one combination
divided by average MSE for the other combination.

For instance with the design RDS, the parabola for the current
estimator (dashed line in Figure 3) is 2.4 times as high as the
parabola for the new estimator.  Equivalently, because of the simple
nature of the ratio estimator, the average MSE for the 13 attribute
variables with the current estimator is 2.4 times the average MSE with
the new estimators.  So the overall relative efficiency of the new
estimator is 2.4.  With the SB design, the overall relative efficiency
of the new estimator is 3.8.  So we get a substantial gain in
efficiency with the new estimators even for variables that, unlike the
link-related variables, do not have an obviously strong or consistent
relationship with the network sampling inclusion probabilities.  The
plots indicate that again much of this gain from from a big reduction
of bias with the new estimators.
  
Confidence interval coverage probabilities for each variable for each
of the 15 variables are given in the Supplemental Tables.  While the
nominal coverage is 95 percent, the actual coverage probability is
assessed in the simulation as the proportion of the 1000 runs,
corresponding to 1000 original samples of size 1200 each, for which
the sample confidence interval contained the true value for the
population.  For the values rounded to 1.0 in the last row of the
table, the exact coverage proportions were 0.999 and 0.998.  The
median coverage probability of the intervals for the 15 characteristics
estimated with each of two designs is .94.


\section{Discussion}

The new estimators obtain better estimates of key population
characteristics from network survey data because the method uses the
full network topology of the data in interaction with the type of
sampling design actually used, which includes branching and
without-replacement sampling.  This is in contrast to currently used
estimators which derive estimates from unrealistic assumptions about
the design and the population network.  Because of the strategic
importance of network surveys for understanding and reducing the HIV
pandemic and emerging epidemics, it is highly desirable and urgent to
bring the new estimators into practice.

The advantage of the new estimators is especially great
for estimating link-related quantities such as mean degree and 
concurrency, for which the values are strongly related to survey
inclusion probabilities.  The link-related variables are
directly related to the network risk of HIV spread in the population.
The elimination of most of the bias through the new estimation
method enables confidence intervals of modest width to have coverage
probabilities close to the desired nominal value.

The new methods work for the freely branching network designs such as
snowball sampling as well as for the limited-branching designs such as
the RDS designs as currently practiced.  The current estimators
perform poorly for the freely branching designs because those designs
are the most far from the assumed random walk.  Computationally the
calculations of the new estimators are fast and scale up well because
of the use of a sampling process approach that stays close to the
desired equilibrium distribution.

The new estimation approach should work also for a wide range of
network related sampling designs and applications.  Because the new
estimation method scales up well, these include studies of online
social network communities and their characteristics and other large
network estimation problems [\cite{papagelis2011sampling},
  \cite{gabielkov2014sampling}].  Another type of network design in
which the new estimators could contribute involves contact tracing for
intervention in outbreaks of sexually transmitted diseases. In these
network designs only contacts of people who test positive are traced
and then tested and treated in necessary.  \cite{peters2016hiv} and
\cite{campbell2017detailed} report on a study in which contract
tracing was used after an HIV outbreak associated with opioid misuse
in Indiana had spread rapidly. The traced network together with
phylogenetic data from sequencing of virus strains was used to
determine where the outbreak started and how it spread.  Because the
contact tracing procedure is another network sampling design
variation, it is possible that the estimators introduced here could
contribute to such network analyses.  Adaptive spatial
designs used for monitoring surveys of rare and endangered plants and
animals can be recast as network sampling problems
(\cite{thompson2006aws}, \cite{thompson2011adaptive}).  Once in that
framework, the inference methods proposed here apply immediately.

One finding that emerges from this study is that the network survey
\textit{designs} as currently used are good and provide invaluable
information. Increased accuracy of \textit{estimates} from these
designs will increase the value of the data collected. The designs are
highly effective at reaching into the key areas of hard-to-reach
populations.  Researchers can feel free to use a wider range of
designs, such as those that more freely branch, as suits each
particular situation.  The new estimates work well with each of these
design variations.  The findings from these surveys are needed for
effective interventions and policy to alleviate critical problems for
vulnerable populations.  The work of policy makers involved with
programs concerned with vulnerable populations will benefit from the
increased value of the survey network data.




\section*{Supplement to New estimates for network sampling }


\section*{Introduction to supplement}

The Supplement includes Details on methods, Supplemental Tables.  The Methods details  section includes additional explanation of the
new estimation method for network surveys and why it works; additional
description of the two approaches, repeated samples and sampling
process, to calculating the inclusion
frequencies $f_i$ which estimate the real-world inclusion
probabilities $p_i$; description of the simulations; additional
details and variations on the estimators and variance estimators; and
location of the data and the source C code for the sampling process algorithm for
estimating the inclusion probabilities.  The Supplemental Tables
include the numbers behind the figures and additional values such as
expected values of the variances of estimators and mean confidence
interval half-widths.

\section*{Details on methods}
 
\subsection*{Repeated samples and sampling process}

The inclusion frequencies $f_i$ are calculated by re-sampling the
sample network many times using a design similar to the original
design used to select the members of the hidden population from the
real world.  Two approaches to the resampling are to repeatedly select
independent resamples, each from seeds to target sample size, and to
select the sequences of resamples using a Markov-chain resampling
process.  


In the simulations of the paper only the sampling process approach was
used.  The repeated-sample approach is illustrated here first as
understanding that makes the sampling-process approach easier to
understand.

Figure 1 of the text shows an RDS sample of 1200 people selected from
the Colorado Springs network population of sex workers, drug users,
and partners of each (\cite{potterat1999network}).  
The target sample size of the resamples is 400.

The simulation study selected 1000 samples of 1200 from the
population, for each of the two designs RDS and SB.  The
simulation study used the sampling process method, which is
computationally very much faster than the independently repeated
samples.

Given the network sample obtained from the real world
network sampling design, we obtain a sequence of re-samples  
\[
\{S_1, S_2, S_3, ...,S_T \}
\] 
from the network data using a fast-sampling process 
similar to the original design.  $T$ is the number of iterations.

For unit $i \in U_s$, there is a sequence of indicator random variables:
\[
\{Z_{i1}, Z_{i2}, Z_{i3},..., Z_{iT} \}
\]
where $ Z_{it} = 1$ if $i\in S_t $ and  $Z_{it} = 0$ if $ i \notin
S_t$, for $t = 1, 2, ..., T$, the number of iterations of the sampling
process.  

The average 
\[
f_i = \frac{1}{T} \sum_{i = 1}^T Z_i
\]
is used as an estimate of the relative inclusion probability  of unit
$i$ in the similar design used to obtain the data from the real
world.  If the real-world network design is done without replacement,
then the fast-sampling process is also carried out without
replacement.  

In the repeated-sample approach, each sample in the sequence proceeds from
selection of seeds to target sample size.  With this approach the
samples in the sequence $\{S_1, S_2, S_3, ...,S_T \}$ are independent
of each other.  

In the sampling-process approach, each sample $S_t$ is selected
dependent on the one before it, $S_{t-1}$.  To get from $S_t$ to
$S_{t-1}$ we probabilistically trace links out from $S_t$, randomly
drop some nodes from $S_t$, and may with low probability select one or
more new seeds.  Advantages of the sampling-process approach are
first, that the computation can be made very fast.  Second, the
sampling process is fast-mixing and once it reaches it's stationary
distribution every subsequent sample $S_t$ is in that distribution.
The stationary distribution of the sequence of samples represents a
balance between the re-seeding distribution, which can be kept small
with a low rate of re-seeding, and the design tendencies arising from
the link-tracing and the without-replacement nature of the selections.

The sampling process is without-replacement in that a node in the
sample can not be selected again while it is in the sample.  Once it
has been removed from the sample it can be selected anew at any time.  With
the sampling-process design, the sequence of fast samples $S_1, S_2, ...$
forms a Markov chain of sets, with the probability of set $S_t$
depending only on the previous set $S_{t-1}$.  



In the simulations the repeated-sample design selects initial seeds using
Bernoulli sampling.  A low rate of re-seeding is used, mainly to ensure that
the growth of the sample can not get stuck before target sample size
is reached.  A medium rate of link-tracing is used.  Links out
are traced with independent Bernoulli selections.  Because no coupons
were used in the re-sampling, each selected node can continue to
recruit without time limit.

The sampling process can use a high tracing rate because
removals offset the tracing to maintain a stochastic balance around
target sample size.
A relatively high rate of ongoing reseeding rate is used so that the
process does not get locked out of any components.  Whenever the sample goes
above target sample size, the sample is randomly thinned with
removals, with probability of removal set to make expected sample size
back within target at the next step.  All these features make the
process very fast mixing.

Specifically, in the examples we trace the links out from the current
sample $S_t$ independently, each with probability $p$.  Nodes are
removed from the sample independently with probability $q$.  The
removal probability $q$ is set adaptively to be
$q_t = (n_t - n_{\rm target}/n_t$ if $n_t > n_{\rm target}$ and $q_t = 0$ otherwise, so
that sample size fluctuates around its target during iterations.
Sampling is without replacement in that a node in $S_t$ is not
while it remains in fast sample, but it may be reselected
at any time after it is removed from the fast sample.  The re-seeding
rate can be low because the seeds at the beginning of each sample
serve to get the sample into enough components.

In the repeated-samples design, the seeding rate $p_s =
0.0167$; the tracing rate is $p = 0.05$, and the re-seeding rate is
$p_r = 0.001$.  The sampling-process design  uses no
initial seeds, relying on the re-seeding to initialize the process and
bring it quickly into its stationary distribution.  The removal rate
$q$ is set adaptively as described above to keep the process in
fluctuation around its target sample size.  The re-seeding rate is $p_r
= 0.01$.  Even though the re-seeding rate is relatively high, at each
step most added nodes are added by tracing links, because that rate
is so much higher.  The re-seeding serves to keep the process from
getting permanently locked out of any component through removals.

If the real-world survey sampling is done with replacement, one can
use a re-sampling  design that is with replacement.  An
advantage of this is that a target sample size for the fast design can
be used that equals the actual sample size used in obtaining the
data.  However, in most cases the real design is done without
replacement.

Sampling processes of these types are discussed in (26)
\cite{thompson2017adaptive}
for their
potential uses as measures of network exposure of a
node, or a measure of network centrality, or a predictive indicator of
regions of a network where an epidemic might next explode.
Calculation of the statistic $f_i$ for each unit in the network sample
can be used as an index of the network exposure of that unit.  A high value
of $f_i$ indicates the unit has high likelihood of being reached by a
network sample such as ours.  It will also have a relatively high
likelihood of being reached by a virus, such as HIV, that spreads on
the same type of links by a link-tracing process that is broadly
similar.  A given risk behavior will be more risky for a person with
high network exposure.  For a person in a less well connected part of
the network, the same behavior carries lower risk.  Since a purpose of
the surveys is to identify risk characteristics, an index of network
exposure measures another dimension of that risk, beyond the individual
behavior and health measures.  Here, however, are interested in their
usefulness for estimating population characteristics based on
link-tracing network sampling designs.

\subsection*{Estimators}

This supplemental section contains additional detail on estimation
formulas.  It includes estimation when the original design is carried
out with replacement and estimation of a ratio.  

Network sampling designs select units with unequal
probabilities.  With unequal probability sampling designs, sample
means and sample proportions do not provide unbiased estimates of
their corresponding population means and proportions.  

To estimate the mean of variable $y$ with an unequal-probability
sampling design, the generalized unequal probability estimator has the
form 
\begin{equation}
\hat\mu_{\pi} = \frac{\sum_s (y_i/\pi_i)}{\sum_s (1/\pi_i)}
\end{equation}
where
$\pi_i$ is inclusion probability of unit $i$.

With the network sampling designs of interest here, the inclusion
probabilities $pi_i$ are not known and can not be calculated from the
sample data.  To circumvent this problem the Volz-Heckathorn Estimator
uses degree, or self-reported number of partners,  to approximate inclusion probability:
\begin{equation}
\hat\mu_{d} = \frac{\sum_s (y_i/d_i)}{\sum_s (1/d_i)}
\end{equation}
in which $d_i$ is the degree, the number of self-reported partners, of person $i$.

The rationale for this approximation is that if the sampling design is
a random walk with replacement, or several independent random walks with replacement and the population
network is connected, then the selection probabilities of the random
walk design will converge over time to be proportional to the $d_i$.
Here connected means that each node in the population can be reached
from any other node by some path, or chain of links, so that the
population network consists of only one connected component.  Biases
in this estimator result from the use of without-replacement sampling
in the real-world designs, the use of coupon numbers $k$ greater than
1 making the design different from a random walk, population networks
being not connected into a single component, or slow mixing due to
specifics of the
population network structure.    

The  new estimator, with a non-replacement
sampling design, is 
\begin{equation}
\hat\mu_f = \frac{\sum_{i \in s} (y_i/f_i)}{\sum_{i \in s} (1/f_i)}
\end{equation}
where
$f_i$ is the inclusion frequency of unit $i$ in the resampling process run
on the sample network data.

A simple variance estimator to go with the new estimator is
\begin{equation}
\widehat{\textrm var}(\hat\mu_f) = \frac{1}{(\sum_{i \in s}
  1/f_i)^2}\sum_{i \in s} \frac{(y_i - \hat\mu_f)^2}{f_i^2}
\label{eq:hatvar1}
\end{equation}

Another simple variance estimator is 
\begin{equation}
\widehat{\textrm var}(\hat\mu) =  \frac{1}{n(n-1)}  \sum_{s}
\left(\frac{ny_i/f_i}{\sum_s (1/f_i)} - \hat\mu_f\right)^2
\label{eq:hatvar2}
\end{equation}

\medskip
\medskip

An approximate $1 - \alpha$ confidence interval is then calculated as 
\begin{equation}
\hat\mu_f \pm z \sqrt{\widehat{\textrm var}(\hat\mu_f)}
\end{equation}
with $z$ the $1-\alpha/2$ quantile from the standard Normal distribution.

The variance estimator \ref{eq:hatvar1} is based on, and simplified from, the Taylor
series linear approximation theory for generalized unequal probability
estimator. 
Linearization leads to the estimator of the variance of the
generalized estimator 
\begin{equation}
\widehat{\textrm var}(\hat\mu_{\pi}) = \frac{1}{(\sum_{i \in s}
  1/\pi_i)^2}\sum_{i \in s} \sum_{j \in s} \check\Delta_{ij}\frac{(y_i
  - \hat\mu_{\pi})}{\pi_i}\frac{(y_j - \hat\mu_{\pi})}{\pi_j}
\end{equation}
where 
\[
\check\Delta_{ij} = \frac{\pi_{ij} - \pi_i\pi_j}{\pi_{ij}}
\]
where $\pi_{ij}$ is the joint inclusion probability for units $i$ and
$j$.  A good discussion of the approach is found in
\cite{sarndal2003model}, with this variance estimator on p. 178 of
that work.  That type of variance estimator goes back to at least to \cite{brewer1983sampling} and is described on p. 178 in \cite{sarndal2003model}.  Here though it has been modified, first to apply to the generalized unequal probability estimatore instead of the Horvitz-Thompson estimator, and second by using the inclusion frequencies $f_i$ to estimate the inclusion probabilities $\pi_i$.  

\medskip

The variance estimator \ref{eq:hatvar2} is based on the idea that if
the sum $\hat\mu_f = \sum_{i \in s} (y_i/f_i)/\sum (1/f_i)$ estimates
$\mu$ then each piece $(y_i/f_i)/\sum (1/f_i)$ estimates $\mu/n$ and
so $t_i = n(y_i/f_i)/\sum (1/f_i)$ would be an estimate of $\mu$, for
$i=1, ..., n$.  Ignoring the dependence from the without-replacement
sampling and treating $t_1, ..., t_n$ as uncorrelated, then
$\hat\mu_f$ is the sample mean of the $t_i$ and \ref{eq:hatvar2} is
their sample variance divided by sample size.  

In simulations both \ref{eq:hatvar1} and \ref{eq:hatvar2} give decent
variance estimates and confidence intervals.  The coverage probability
tended to be modestly better with \ref{eq:hatvar2}, and that is the
one used in the simulations of the report.  

Consider an estimator of the variance using the full variance
expression with  the fast-sample
frequencies $f_i$ in place of the $\pi_i$ and, in place of the joint
inclusion probability $\pi_{ij}$, the frequency $f_{ij}$ of inclusion
of inclusion of both units $i$ and $j$ in the fast sampling process.
This would give 
\begin{equation}
\widehat{\textrm var}(\hat\mu_f) = \frac{1}{(\sum_{i \in s}
  1/f_i)^2}\sum_{i \in s} \sum_{j \in s} \hat\Delta_{ij}\frac{(y_i - \hat\mu_f)}{f_i}\frac{(y_j - \hat\mu_f)}{f_j}
\end{equation}
where 
\[
\hat\Delta_{ij} = \frac{f_{ij} - f_if_j}{f_{ij}}
\]

\medskip

The double sum in the variance estimate expression will have
$n(n-1)/2$ terms in which $i \ne j$.  The most influential of these
terms are the ones in which the joint frequency of inclusion $f_{ij}$
is relatively large.  Because of the link tracing in the fast sampling
process, sample unit pairs with a direct link between them will tend
occur together more frequently than those without a direct link.  An
estimator using only those pairs with known links between them in the
sample data would be 

\begin{equation}
\widehat{\textrm var}(\hat\mu_f) = \frac{1}{(\sum_{i \in s}
  1/f_i)^2} \left(\sum_{i \in U_s} (f_i - 1)\frac{(y_i -
    \hat\mu_f)^2}{f_i}  +  \sum_{(i,j) \in E_s} \hat\Delta_{ij}\frac{(y_i - \hat\mu_f)}{f_i}\frac{(y_j - \hat\mu)}{f_j}\right)
\end{equation}
where $E_s$ is the sample edge set.  That is, $E_s$ consists of the
known edges $(i,j)$ between pairs of units in the sample data.  In
general the size of the sample edge set $E_s$ will be much smaller
that the $n^2$ possible sample node pairings $(i, j)$, or the
$n(n-1)/2$ pairings with $i\ne j$, where $n$ is the sample size.  

A further simplification and approximation for estimating the variance
of the estimator is to use only the diagonal terms, that is, 
\begin{equation}
\widehat{\textrm var}(\hat\mu_f) = \frac{1}{(\sum_{i \in s}
  1/f_i)^2} \sum_{i \in s} (1 - f_i)\frac{(y_i -
    \hat\mu_f)^2}{f_i} 
\end{equation}

Dropping the coefficients $(1-f_i)$, each of which is less than or
equal to one, gives an estimate of variance that is larger, leading to
wider, more conservative confidence intervals.

If the real-world network sampling design and correspondingly the 
re-sampling process are  with-replacement, the estimator of $\mu$ is  
\begin{equation}
\hat\mu_f =  \frac{\sum_{i \in s}
    (m_{i}y_i/g_i)}{\sum_{i \in s}
(m_{i}/g_i)}
\end{equation}
in which $m_i$ is the number of times unit i is selected in the real
design and $g_i$ is the average number of selection counts of unit
$i$ in the fast sampling process.

If the real-world sampling design is with replacement, the re-sampling
process can be done with replacement.  In that case  let $m_t(i)$ be the number of times
node $i$ is selected at iteration $t$.  The quantity
$g_i = (1/t) \sum_{s=1}^t m_t(i)$, the average number of selections up to
iteration $t$, estimates the expected number of selections for node
$i$ under the with-replacement design at any given iteration $t$.

With a with-replacement fast design the corresponding variance
estimator is
\begin{equation}
\widehat{\textrm var}(\hat\mu_f) =   \frac{1}{(\sum_{i \in s}
  m_{i}/g_i)^2}\sum_{i \in s} \frac{m_{i}(y_i - \hat\mu_f)^2}{g_i^2}
\end{equation}

If $x_i$ is another variable, an estimator of the ratio
$R=\mu_y/\mu_x$ of the mean of
$y$ to the mean of $x$ is
\begin{equation}
\hat R = \frac{\sum_{i \in s} y_i/f_i}{\sum_{i\in s} x_i/f_i}
\end{equation}
with simple variance estimator
\begin{equation}
\widehat{\textrm var}(\hat R) = \frac{1}{(\sum_{i \in s}
  x_i/f_i)^2}\sum_{i \in s} \frac{(y_i - x_i\hat\mu_f)^2}{f_i^2}
\end{equation}

\subsection*{Data}

The new and current estimators were evaluated using the  entire network data set of 5492
people and 21,644 links from the Colorado Springs Project 90 study on
the heterosexual spread of HIV
\cite{potterat1999network}.  The study protocol was approved by the Human Subjects Committee of Colorado Health Sciences Center and included written informed consent (41) \cite{woodhouse1994mapping}.
The 
data are available to researchers
(https://opr.princeton.edu/archive/p90/).  The links combine drug,
sexual, and social relationships.  The study was so thorough in tracing every relationship link that it is the closest we have to data on an entire at-risk hidden network population of the kind in which we are interested.   The same
data set was used in simulations in \cite {goel2010assessing},\cite{baraff2016estimating}, and \cite{fellows2018respondent}.  The other studies used only the largest connected component of 4430 people, possibly because of the assumption of a single-component network required along with the random walk design assumption used in justifying the current estimators.  That leaves out 1062 people in smaller components.  The Colorado Springs at-risk population is consistent with many other real-world networks in having one very large component and a number of smaller components.  Therefore we have used the full network population for realism in the simulations.

\subsection*{Simulations}

For each of the two designs, 1000 samples of target size $n = 1200$
were selected from the 5492 study population.  In RDS, 3
coupons were given to each respondent (fewer if the respondent had
fewer than 3 partners).  In SB , the coupon maximum was 15.
Coupons had an expiration date 28 days from issue.  Seeds (240 or 20\%
of $n$) were selected at random.  The resampling process, like
the original design, used link-tracing, branching, and
without-replacement sampling.   No  coupons were used in the resampling process,
so that the same resampling design was used for each of the four
original designs.  As can
be seen from the RDS sample in Figure 1 where each
respondent was given no more than 3 coupons with which to, a
without-replacement resample can at no point branch more than 3 in any
case, or up to 4 branches from a re-seed.  For each of the 1000 samples, the new estimator was
calculated by selecting $T=10,000$ resamples each of target size 400
and averaging the inclusion indicators for each of the 1200 sample
people, giving the frequencies $f_i$ to calculate the estimate
$\hat\mu_f$.

\section*{Supplemental Tables}

Tables S1-S4 give the numbers behind the figures in the paper.  In
addition to the 13 attribute variables in the node data file of the
Colorado Springs Project 90 data, the tables include two variables
whose values are calculated from the link data file.  These are
degree, the number of partners a person has, and ``deg2plus'', an
indicator of whether the person has two or more partners.  The
population proportion of people with two or more partners, which is the mean of the indicator variable deg2plus,  is also
referred to as concurrency.  

The column ``actual'' gives the population mean or proportion for each
variable.  ``E.est'' is the mean value of the estimator.  Bias is
E.est - actual.  The standard deviation ``sd'' is $\sqrt{{\rm
var}(\hat\mu)}$ for the given estimator.  The mean squared error
``mse'' is ${\rm E}((\hat\mu -\mu)^2)$.  The relative efficiency
``eff'' is ${\rm E}((\hat\mu_d -\mu)^2)/{\rm E}((\hat\mu_f -\mu)^2)$
for the current estimator $\hat\mu_d$.  For the sample mean $\bar y$
the relative efficiency is similarly defined with the mean squared
error of $\bar y$ in the numerator and that of the new estimator
in the denominator.  The relative bias is the ratio of absolute
biases, with the bias of the new estimator in the denominator.

Tables S5-S8 give confidence interval coverage probability for nominal
95 percent confidence intervals.  They expand on the information in
the text by giving the average half-width of the interval for each
variable.  Since the intervals are of the symmetric form estimate
$\pm$ half-width of interval, it is natural to look at the average
half-width in relation to the actual value of what is being
estimated.  Coverage probability is the proportion of simulation runs
for which the interval covers the true value.  


\begin{table}[ht]
\centering
\caption{RDS Table} 
\begin{tabular}{lrrrrrrr}
  \hline
New & actual & E.est & bias & sd & mse & eff & rbias \\ 
  \hline
degree & 7.88 & 8.21 & 0.327829 & 0.320172 & 0.209982 & 1.00 & 1.00 \\ 
  nonwhite & 0.24 & 0.25 & 0.010668 & 0.021759 & 0.000587 & 1.00 & 1.00 \\ 
  female & 0.43 & 0.43 & 0.001866 & 0.022826 & 0.000525 & 1.00 & 1.00 \\ 
  worker & 0.05 & 0.06 & 0.003409 & 0.009484 & 0.000102 & 1.00 & 1.00 \\ 
  procurer & 0.02 & 0.02 & 0.001456 & 0.004622 & 0.000023 & 1.00 & 1.00 \\ 
  client & 0.09 & 0.09 & 0.000230 & 0.015043 & 0.000226 & 1.00 & 1.00 \\ 
  dealer & 0.06 & 0.07 & 0.005925 & 0.010702 & 0.000150 & 1.00 & 1.00 \\ 
  cook & 0.01 & 0.01 & -0.000019 & 0.003939 & 0.000016 & 1.00 & 1.00 \\ 
  thief & 0.02 & 0.02 & 0.001522 & 0.006084 & 0.000039 & 1.00 & 1.00 \\ 
  retired & 0.03 & 0.03 & 0.000644 & 0.008213 & 0.000068 & 1.00 & 1.00 \\ 
  homemakr & 0.06 & 0.06 & -0.000079 & 0.010703 & 0.000115 & 1.00 & 1.00 \\ 
  disabled & 0.04 & 0.04 & 0.001477 & 0.009087 & 0.000085 & 1.00 & 1.00 \\ 
  unemploy & 0.16 & 0.17 & 0.006565 & 0.016631 & 0.000320 & 1.00 & 1.00 \\ 
  homeless & 0.01 & 0.01 & 0.000673 & 0.004938 & 0.000025 & 1.00 & 1.00 \\ 
  deg2plus & 0.82 & 0.82 & 0.003064 & 0.021839 & 0.000486 & 1.00 & 1.00 \\ 
    \hline
Current & actual & E.est & bias & sd & mse & eff & rbias \\ 
  \hline
  degree & 7.88 & 5.44 & -2.447003 & 0.215896 & 6.034435 & 28.74 & 7.46 \\ 
  nonwhite & 0.24 & 0.26 & 0.021276 & 0.021866 & 0.000931 & 1.58 & 1.99 \\ 
  female & 0.43 & 0.41 & -0.023358 & 0.022209 & 0.001039 & 1.98 & 12.52 \\ 
  worker & 0.05 & 0.05 & -0.004432 & 0.009785 & 0.000115 & 1.14 & 1.30 \\ 
  procurer & 0.02 & 0.01 & -0.003378 & 0.003509 & 0.000024 & 1.01 & 2.32 \\ 
  client & 0.09 & 0.13 & 0.038526 & 0.017978 & 0.001808 & 7.99 & 167.37 \\ 
  dealer & 0.06 & 0.06 & 0.001224 & 0.010642 & 0.000115 & 0.77 & 0.21 \\ 
  cook & 0.01 & 0.01 & -0.001382 & 0.002870 & 0.000010 & 0.65 & 71.11 \\ 
  thief & 0.02 & 0.02 & -0.000867 & 0.005624 & 0.000032 & 0.82 & 0.57 \\ 
  retired & 0.03 & 0.03 & 0.003194 & 0.008281 & 0.000079 & 1.16 & 4.96 \\ 
  homemakr & 0.06 & 0.05 & -0.008379 & 0.008309 & 0.000139 & 1.22 & 105.95 \\ 
  disabled & 0.04 & 0.04 & -0.004888 & 0.007583 & 0.000081 & 0.96 & 3.31 \\ 
  unemploy & 0.16 & 0.13 & -0.028841 & 0.013125 & 0.001004 & 3.14 & 4.39 \\ 
  homeless & 0.01 & 0.01 & -0.000774 & 0.004193 & 0.000018 & 0.73 & 1.15 \\ 
  deg2plus & 0.82 & 0.64 & -0.184948 & 0.027210 & 0.034946 & 71.86 & 60.35 \\ 
    \hline
$\bar y$ & actual & E.est & bias & sd & mse & eff & rbias \\ 
  \hline
  degree & 7.88 & 14.32 & 6.435291 & 0.235165 & 41.468275 & 197.48 & 19.63 \\ 
  nonwhite & 0.24 & 0.28 & 0.040718 & 0.017822 & 0.001976 & 3.36 & 3.82 \\ 
  female & 0.43 & 0.47 & 0.033170 & 0.011699 & 0.001237 & 2.36 & 17.78 \\ 
  worker & 0.05 & 0.09 & 0.041124 & 0.006060 & 0.001728 & 17.01 & 12.06 \\ 
  procurer & 0.02 & 0.03 & 0.015914 & 0.003251 & 0.000264 & 11.24 & 10.93 \\ 
  client & 0.09 & 0.07 & -0.014696 & 0.007013 & 0.000265 & 1.17 & 63.84 \\ 
  dealer & 0.06 & 0.12 & 0.054420 & 0.006879 & 0.003009 & 20.11 & 9.19 \\ 
  cook & 0.01 & 0.01 & 0.001495 & 0.002406 & 0.000008 & 0.52 & 76.94 \\ 
  thief & 0.02 & 0.04 & 0.014992 & 0.003999 & 0.000241 & 6.12 & 9.85 \\ 
  retired & 0.03 & 0.03 & -0.000374 & 0.004017 & 0.000016 & 0.24 & 0.58 \\ 
  homemakr & 0.06 & 0.07 & 0.007123 & 0.005985 & 0.000087 & 0.76 & 90.07 \\ 
  disabled & 0.04 & 0.06 & 0.014588 & 0.005215 & 0.000240 & 2.83 & 9.87 \\ 
  unemploy & 0.16 & 0.25 & 0.090174 & 0.010090 & 0.008233 & 25.75 & 13.73 \\ 
  homeless & 0.01 & 0.02 & 0.003934 & 0.002662 & 0.000023 & 0.91 & 5.85 \\ 
  deg2plus & 0.82 & 0.93 & 0.110389 & 0.007364 & 0.012240 & 25.17 & 36.02 \\ 
   \hline
\end{tabular}
\label{rdstable}
\end{table}

\begin{table}[ht]
\centering
\caption{SB Table}
\begin{tabular}{lrrrrrrr}
  \hline
New & actual & E.est & bias & sd & mse & eff & rbias \\ 
  \hline
degree & 7.88 & 8.30 & 0.415435 & 0.274937 & 0.248177 & 1.00 & 1.00 \\ 
  nonwhite & 0.24 & 0.26 & 0.017506 & 0.021710 & 0.000778 & 1.00 & 1.00 \\ 
  female & 0.43 & 0.43 & 0.000860 & 0.022582 & 0.000511 & 1.00 & 1.00 \\ 
  worker & 0.05 & 0.06 & 0.006121 & 0.009902 & 0.000136 & 1.00 & 1.00 \\ 
  procurer & 0.02 & 0.02 & 0.002969 & 0.004545 & 0.000029 & 1.00 & 1.00 \\ 
  client & 0.09 & 0.09 & 0.004574 & 0.014924 & 0.000244 & 1.00 & 1.00 \\ 
  dealer & 0.06 & 0.07 & 0.009554 & 0.010335 & 0.000198 & 1.00 & 1.00 \\ 
  cook & 0.01 & 0.01 & 0.000158 & 0.003987 & 0.000016 & 1.00 & 1.00 \\ 
  thief & 0.02 & 0.02 & 0.002823 & 0.006376 & 0.000049 & 1.00 & 1.00 \\ 
  retired & 0.03 & 0.03 & 0.000947 & 0.007986 & 0.000065 & 1.00 & 1.00 \\ 
  homemakr & 0.06 & 0.06 & -0.001269 & 0.010867 & 0.000120 & 1.00 & 1.00 \\ 
  disabled & 0.04 & 0.04 & 0.001743 & 0.008995 & 0.000084 & 1.00 & 1.00 \\ 
  unemploy & 0.16 & 0.17 & 0.009528 & 0.015928 & 0.000344 & 1.00 & 1.00 \\ 
  homeless & 0.01 & 0.01 & 0.000704 & 0.004722 & 0.000023 & 1.00 & 1.00 \\ 
  deg2plus & 0.82 & 0.83 & 0.005025 & 0.021135 & 0.000472 & 1.00 & 1.00 \\ 
    \hline
Current & actual & E.est & bias & sd & mse & eff & rbias \\ 
  \hline
  degree & 7.88 & 5.20 & -2.679280 & 0.198895 & 7.218099 & 29.08 & 6.45 \\ 
  nonwhite & 0.24 & 0.27 & 0.032360 & 0.022007 & 0.001531 & 1.97 & 1.85 \\ 
  female & 0.43 & 0.39 & -0.038747 & 0.021683 & 0.001971 & 3.86 & 45.07 \\ 
  worker & 0.05 & 0.05 & -0.001975 & 0.009839 & 0.000101 & 0.74 & 0.32 \\ 
  procurer & 0.02 & 0.01 & -0.002209 & 0.003401 & 0.000016 & 0.56 & 0.74 \\ 
  client & 0.09 & 0.15 & 0.063940 & 0.019839 & 0.004482 & 18.40 & 13.98 \\ 
  dealer & 0.06 & 0.07 & 0.008865 & 0.010872 & 0.000197 & 0.99 & 0.93 \\ 
  cook & 0.01 & 0.01 & -0.001520 & 0.002682 & 0.000010 & 0.60 & 9.62 \\ 
  thief & 0.02 & 0.02 & 0.002340 & 0.006623 & 0.000049 & 1.01 & 0.83 \\ 
  retired & 0.03 & 0.03 & 0.004822 & 0.008194 & 0.000090 & 1.40 & 5.09 \\ 
  homemakr & 0.06 & 0.05 & -0.012629 & 0.008258 & 0.000228 & 1.90 & 9.95 \\ 
  disabled & 0.04 & 0.04 & -0.005893 & 0.007172 & 0.000086 & 1.03 & 3.38 \\ 
  unemploy & 0.16 & 0.13 & -0.030013 & 0.012548 & 0.001058 & 3.07 & 3.15 \\ 
  homeless & 0.01 & 0.01 & -0.000593 & 0.004002 & 0.000016 & 0.72 & 0.84 \\ 
  deg2plus & 0.82 & 0.62 & -0.206160 & 0.027626 & 0.043265 & 91.68 & 41.02 \\ 
    \hline
$\bar y$ & actual & E.est & bias & sd & mse & eff & rbias \\ 
  \hline
  degree & 7.88 & 14.24 & 6.359845 & 0.203947 & 40.489220 & 163.15 & 15.31 \\ 
  nonwhite & 0.24 & 0.30 & 0.057690 & 0.016618 & 0.003604 & 4.63 & 3.30 \\ 
  female & 0.43 & 0.46 & 0.025993 & 0.011495 & 0.000808 & 1.58 & 30.24 \\ 
  worker & 0.05 & 0.10 & 0.047894 & 0.005763 & 0.002327 & 17.17 & 7.82 \\ 
  procurer & 0.02 & 0.03 & 0.019219 & 0.003017 & 0.000378 & 12.84 & 6.47 \\ 
  client & 0.09 & 0.09 & -0.000762 & 0.007672 & 0.000059 & 0.24 & 0.17 \\ 
  dealer & 0.06 & 0.13 & 0.062733 & 0.006721 & 0.003981 & 20.09 & 6.57 \\ 
  cook & 0.01 & 0.01 & 0.001608 & 0.002173 & 0.000007 & 0.46 & 10.18 \\ 
  thief & 0.02 & 0.04 & 0.017587 & 0.004122 & 0.000326 & 6.71 & 6.23 \\ 
  retired & 0.03 & 0.03 & 0.000806 & 0.003848 & 0.000015 & 0.24 & 0.85 \\ 
  homemakr & 0.06 & 0.06 & 0.003532 & 0.006186 & 0.000051 & 0.42 & 2.78 \\ 
  disabled & 0.04 & 0.06 & 0.015092 & 0.005256 & 0.000255 & 3.04 & 8.66 \\ 
  unemploy & 0.16 & 0.26 & 0.094729 & 0.009813 & 0.009070 & 26.33 & 9.94 \\ 
  homeless & 0.01 & 0.02 & 0.004726 & 0.002590 & 0.000029 & 1.27 & 6.71 \\ 
  deg2plus & 0.82 & 0.93 & 0.103386 & 0.007816 & 0.010750 & 22.78 & 20.57 \\ 
   \hline
\end{tabular} 
\label{sbtable}
\end{table}


\begin{table}[ht]
\centering
\caption{RDS: Confidence Interval Coverage} 
\begin{tabular}{lrrr}
  \hline
name & actual & halfwidth & coverage \\ 
  \hline
degree & 7.88 & 0.55 & 0.77 \\ 
  nonwhite & 0.24 & 0.04 & 0.95 \\ 
  female & 0.43 & 0.06 & 0.98 \\ 
  worker & 0.05 & 0.02 & 0.95 \\ 
  procurer & 0.02 & 0.01 & 0.90 \\ 
  client & 0.09 & 0.03 & 0.95 \\ 
  dealer & 0.06 & 0.02 & 0.95 \\ 
  cook & 0.01 & 0.01 & 0.78 \\ 
  thief & 0.02 & 0.01 & 0.94 \\ 
  retired & 0.03 & 0.02 & 0.93 \\ 
  homemakr & 0.06 & 0.02 & 0.94 \\ 
  disabled & 0.04 & 0.02 & 0.94 \\ 
  unemploy & 0.16 & 0.03 & 0.95 \\ 
  homeless & 0.01 & 0.01 & 0.88 \\ 
  deg2plus & 0.82 & 0.07 & 1.00 \\ 
   \hline
\end{tabular}
\label{RDScoverage}
\end{table}

\begin{table}[ht]
\centering
\caption{SB: Confidence Interval Coverage} 
\begin{tabular}{lrrr}
  \hline
name & actual & halfwidth & coverage \\ 
  \hline
degree & 7.88 & 0.56 & 0.72 \\ 
  nonwhite & 0.24 & 0.04 & 0.92 \\ 
  female & 0.43 & 0.06 & 0.99 \\ 
  worker & 0.05 & 0.02 & 0.95 \\ 
  procurer & 0.02 & 0.01 & 0.95 \\ 
  client & 0.09 & 0.03 & 0.96 \\ 
  dealer & 0.06 & 0.02 & 0.94 \\ 
  cook & 0.01 & 0.01 & 0.79 \\ 
  thief & 0.02 & 0.01 & 0.93 \\ 
  retired & 0.03 & 0.02 & 0.93 \\ 
  homemakr & 0.06 & 0.02 & 0.93 \\ 
  disabled & 0.04 & 0.02 & 0.94 \\ 
  unemploy & 0.16 & 0.03 & 0.95 \\ 
  homeless & 0.01 & 0.01 & 0.86 \\ 
  deg2plus & 0.82 & 0.07 & 1.00 \\ 
   \hline
\end{tabular}
\label{SBcoverage}
\end{table}

\section*{Acknowledgements}
This research was supported by Natural Science and Engineering
Research Council of Canada (NSERC) Discovery grant RGPIN327306.  I
would like to thank John Potterat and Steve Muth for making the
Colorado Springs study data available  and for their generous help 
explaining it. I would like to express appreciation for the
participants in that study who shared their personal information with
the researchers so that it could be made available in anonymized form
to the research community and contribute to a solution to HIV and
addiction epidemics and to basic understanding of social networks.


\bibliography{fastrefs}

\end{document}